\documentclass{stsci_report}

\usepackage{graphicx}
\usepackage{siunitx}
\usepackage{todonotes}
\usepackage{pdfpages}
\usepackage{indentfirst}
\usepackage{todonotes}
\usepackage[breaklinks]{hyperref} 
\usepackage[hyphenbreaks]{breakurl}

\copyrighttext{Copyright\copyright\ \the\year\ The Association of Universities for Research in Astronomy, Inc. All Rights Reserved.}

\presubtitle{Instrument Science Report WFC3 2025-01}
\title{Analyzing Exoplanet Transits Observed with the WFC3/UVIS G280 Grism}
\author{Munazza K. Alam, Frederick Dauphin, Amanda Pagul}
\date{April 21, 2025}

\begin{document}

\maketitle

\abstract{Here we describe a Jupyter notebook demonstrating methods for the reduction and analysis of exoplanet transit observations taken with the WFC3/UVIS G280 grism. Released on
Space Telescope's $\tt {hst\_ notebooks}$ GitHub repository, this notebook presents an example workflow for processing time-series observations taken with the G280 grism -- from the calibrated flat-fielded spectra to transit light curves ready for fitting. The specific routines presented in the notebook are explained here, and are meant to highlight data reduction steps that users will typically apply to extract transit light curves. The steps include background subtraction, spatial and temporal cosmic ray correction, spectral trace fitting, spectral extraction, and light curve generation. The end products of the routines in the Jupyter notebook are the raw broadband and spectroscopic light curves, which can be ingested into publicly available light curve fitting tools to extract planetary transmission spectra.}

\section{Introduction}
\label{sec:intro}

Time-series observations of transiting exoplanets can allow us to measure their atmospheric compositions. The Hubble Space Telescope has paved the way for such observations over the past two decades, with the Space Telescope Imaging Spectrograph (STIS) G430L \& G750L ($\sim$0.3--1\,$\mu$m) and the Wide Field Camera 3 (WFC3) G102 \& G141 ($\sim$0.8--1.7\,$\mu$m) serving as the workhorse modes for transit spectroscopy. Recently, however, WFC3's G280 grism (0.2--0.8\,$\mu$m) has become popular among the exoplanet community thanks to recent studies demonstrating its usefulness for transit observations (e.g., \textcite{wakeford2020,lothringer2022,boehm2024}). In particular, G280's UV-optical wavelength range can yield critical information about the cloud and haze properties, UV absorbers, and photochemistry in exoplanet atmospheres (e.g., \textcite{gao2021,zahnle2009,tsai2021}). Compared to STIS, the WFC3/UVIS G280 grism yields higher precision exoplanet spectra at UV-optical wavelengths with a \textit{single} transit (\textcite{wakeford2020}).  
With over 20 planets observed with the G280 grism since 2020 (in addition to over 100 planets observed with the G102 and G141 grisms since 2012), WFC3 remains the most popular Hubble instrument for observations of transiting exoplanets.  

Despite G280's increased use for exoplanet transit observations, users may face several challenges in reducing G280 grism data, including: a strongly curved spectral trace at bluer wavelengths, a non-linear wavelength solution that must be extracted relative to the spectral trace position, multiple spectral orders overlapping at wavelengths redder than 0.55 $\mu$m for both the positive and negative orders (see Section 8.2 in 
the WFC3 Instrument Handbook\footnote{WFC3 Instrument Handbook: \url{https://hst-docs.stsci.edu/wfc3ihb}}), and an increased number of cosmic rays due to the grism's wide wavelength coverage. To lower the barrier to entry for users interested in analyzing G280 observations, more publicly available tools and resources outlining data reduction prescriptions are needed. The intended audience for this tutorial notebook is users with little to no experience working with G280 and/or exoplanet transit data. 

In this report, we describe a Jupyter notebook demonstrating the reduction and analysis of time-series transit observations taken with the WFC3/UVIS G280 grism. The structure of the report is as follows: in Section \ref{sec:obs}, we describe the observations used in the tutorial notebook. In Section \ref{sec:workflow}, we outline the general notebook set up and workflow. We detail all of the routines used in the data reduction in Section \ref{sec:data_reduction}. Finally, we summarize in Section \ref{sec:summary}.

\section{Observations}
\label{sec:obs}

In the Jupyter notebook presented here, we walk through the reduction of a transit observation of the giant exoplanet HAT-P-41b (PID: 15288; \textcite{wakeford2020}), which was the first time-series transit observed with the WFC3/UVIS G280 grism. Here we focus on the first of the two visits observed in this program. The transit was observed on UT 01 August 2018 and consisted of 54 exposures over five HST orbits, with exposure times of 190 seconds each. The total visit duration was 7.07 hours. A Y POS TARG of $-50$" was applied to place the spectral trace in the center of chip 2. A 2100 $\times$ 800 subarray was used, which is large enough to include both the $+1$ and $-1$ spectral orders. 

\section{General Workflow}
\label{sec:workflow}

Our Jupyter notebook\footnote{G280 Transits Notebook: \url{https://github.com/spacetelescope/hst_notebooks/tree/main/notebooks/WFC3/uvis_g280_transit}} outlining the data reduction procedure for G280 time-series transit observations is hosted on Space Telescope's {$\tt hst\_notebooks$} GitHub repository under the {$\tt notebooks/WFC3/uvis\_g280\_transit$} directory. The directory includes: 

\begin{enumerate}
    \item {$\tt requirements.txt$}: a requirements file for the versions of Python packages (and their dependencies) needed to run the notebook; 
    \item {$\tt g280\_transit\_tools.py$}: a custom module containing functions specific to the reduction and analysis of time-series transit data taken with the G280 grism; and 
    \item {$\tt G280\_Exoplanet\_Transits.ipynb$}: a Jupyter notebook demonstrating the G280 data reduction for a single transit observation of the hot Jupiter HAT-P-41b using the custom {$\tt g280\_transit\_tools.py$} module.
\end{enumerate}
     
Any issues or bugs encountered in running this notebook should be reported to the HST General Help Desk\footnote{HST Help Desk: \url{https://stsci.service-now.com/hst}}.  

\subsection{Downloading Grism Observations from MAST}
\label{sec:mast}

Prior to beginning the data reduction, we must query the observations required to complete the tutorial from the Mikulski Archive for Space Telescopes (MAST). Since we walk through the reduction of one of the transit observations of the giant exoplanet HAT-P-41b (PID: 15288; \textcite{wakeford2020}), we need to query only the first visit of the proposal and reject bias frames. To do so, we use a wildcard for {$\tt obs\_id$} and specify {$\tt target\_name$}. 

For our time-series analysis, we only need to download the calibrated, flat-fielded exposure {$\tt flt$} files produced using the {$\tt calwf3$}\footnote{calwf3: \url{https://www.stsci.edu/hst/instrumentation/wfc3/software-tools/pipeline}} pipeline. We thus filter for the {$\tt flt$} files only, and download them using {$\tt astroquery.mast. Observations$} (\textcite{Ginsburg19}\footnote{astroquery: \url{https://astroquery.readthedocs.io/}}). We then create a {$\tt direct\_image$} directory for the direct image taken before the start of the grism spectral time-series. We also organize the files in the {$\tt mastDownloads$} folder into a {$\tt data$} directory with a subdirectory called {$\tt flt$}, where we move the calibrated flat-fielded spectra downloaded from MAST. We also create subdirectories called {$\tt flt\_clean$}, and {$\tt flt\_full$} for the cleaned flat-fielded and full-frame flat-fielded files, respectively, which will be populated by the outputs of the Jupyter notebook.  After organizing the relevant files in the appropriate directories for our time-series analysis, we can reduce and analyze the time-series spectra. 

\section{Data Reduction Routines}
\label{sec:data_reduction}

We follow the standard data reduction steps for time-series observations of transiting exoplanets, with some steps specific to G280 grism data. The data reduction routines covered in the Jupyter notebook include: background subtraction (Section \ref{sec:bkg_sub}), spatial and temporal cosmic ray correction (Section \ref{sec:cr_corr}), embedding the G280 subarrays into a full-frame image (Section \ref{sec:embed_sub}), spectral trace fitting (Section \ref{sec:trace_fitting}), extraction of the 1D spectral time-series (Sections \ref{sec:spec_extract} and \ref{sec:time_series}), and generation of transit light curves (Section \ref{sec:lcs}).    

\subsection{Background Subtraction}
\label{sec:bkg_sub}

The first step in the data reduction process is to perform a background subtraction for all of the spectra in the time-series. There are several background subtraction methods, but here we use the publicly available G280 sky frames\footnote{G280 sky frames: \url{https://www.stsci.edu/hst/instrumentation/wfc3/documentation/grism-resources/uvis-grism-sky-images}}, which must be downloaded prior to execution. WFC3/UVIS is comprised of two separate charge coupled devices (CCDs), UVIS1 and UVIS2, so the sky calibration is independent for each chip. The sky frames were generated by stacking all public on-orbit science exposures in MAST (as of July 2023) and aggressively masking sources using a modified segmentation map (\textcite{pagul2023}).    

\begin{figure}[!ht]
\begin{centering}
\includegraphics[trim=2cm 0 0 0,width=0.65\textwidth]{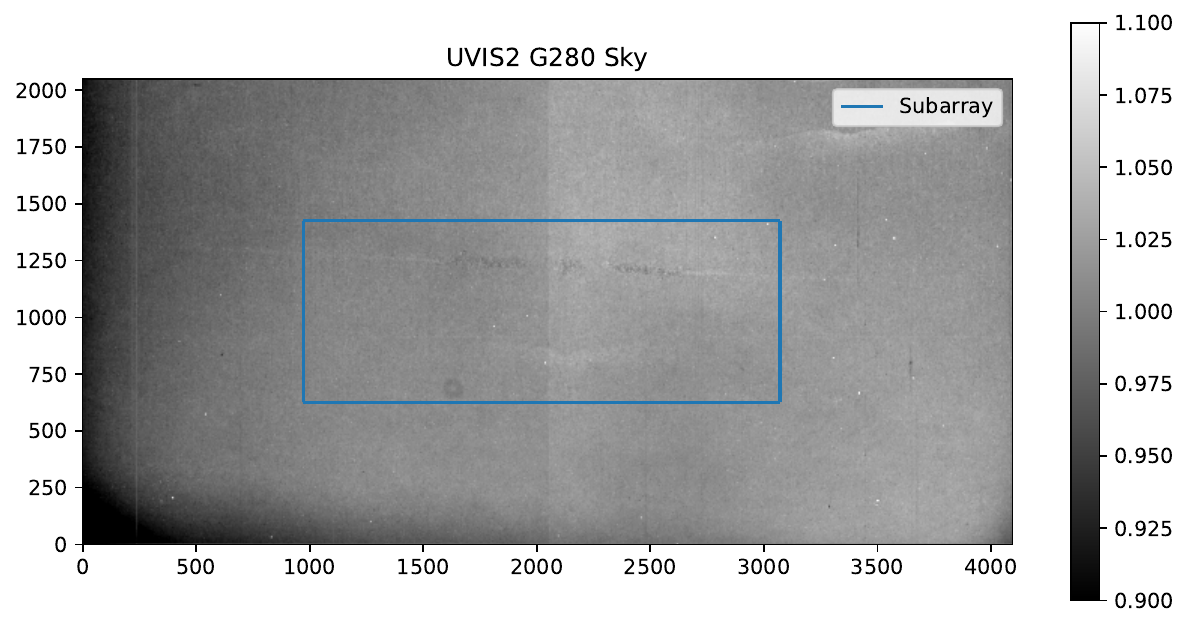}
\caption{The full-frame G280 sky image for UVIS2 with the subarray size and location outlined in blue.} 
\label{fig:g280_sky}   
\end{centering}
\end{figure} 

\begin{figure}[!ht]
\begin{centering}
\includegraphics[trim=0cm 0 0 0cm,width=0.49\textwidth]{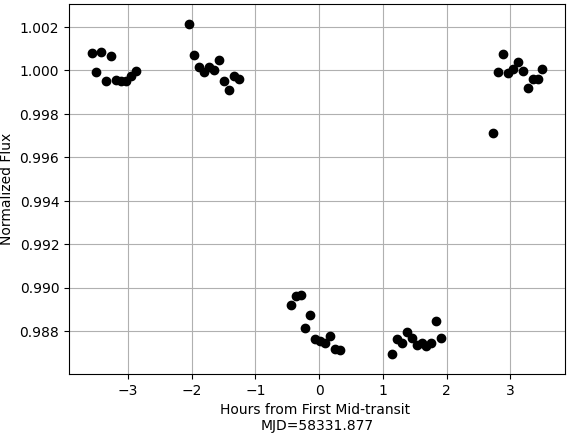}
\includegraphics[trim=0cm 0 0 0cm,width=0.50\textwidth]{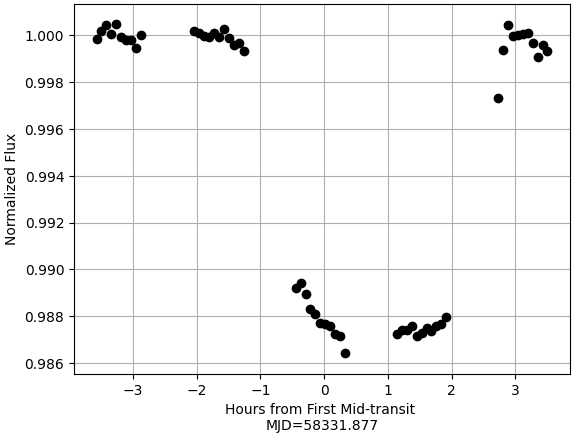}
\caption{The extracted white light curve for HAT-P-41b using a histogram background subtraction method (left) compared to a G280 sky background subtraction (right).} 
\label{fig:bkg_sub}   
\end{centering}
\end{figure}

Since the HAT-P-41b G280 grism data are taken on UVIS2, we load the UVIS2 sky frame as shown in Figure \ref{fig:g280_sky}. We then scale the sky image by the median of each exposure and apply the background subtraction to each science frame in the time-series. Compared to other background subtraction methods in the literature (e.g., the histogram background subtraction of \textcite{wakeford2020}), using the median-stacked sky image provides a more precise extracted white light curve as shown in Figure \ref{fig:bkg_sub}.   

\subsection{Cosmic Ray Correction}
\label{sec:cr_corr}

After applying the background subtraction, we remove cosmic rays, which not only can hit the detector throughout the course of the hours-long time-series observations, but may also be more frequent in G280 data (compared to, e.g., STIS G430L observations) due to its wider wavelength coverage. When recorded on 2D time-series spectra, cosmic rays can artificially increase the flux of the target star in random locations on the detector along the spectra and also appear as outlier data points in the transit light curves in the affected wavelength bins. It is therefore crucial to identify and correct potential cosmic ray hits on 2D spectra both within an exposure (spatially) and also across exposures in the time-series (temporally). A thorough cosmic ray correction can ensure that cosmic ray events do not become a potential source of instrument systematic effects in the transit light curves.   

\begin{figure}[!ht]
\begin{centering}
\includegraphics[width=0.8\textwidth]{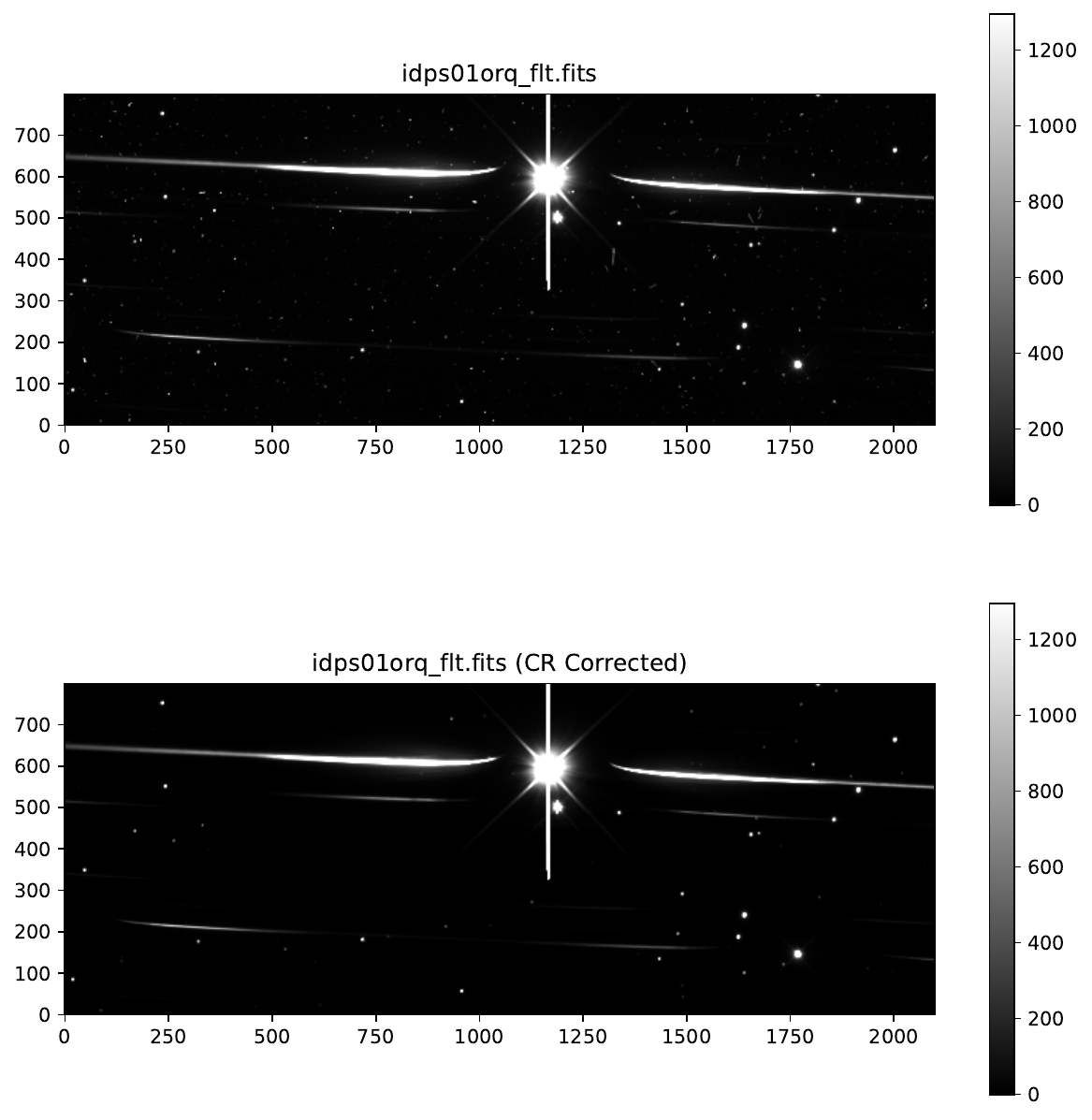}
\caption{The calibrated 2D corrected image (top) compared to the background-subtracted and cosmic-ray-corrected image (bottom).} 
\label{fig:cr_corr}   
\end{centering}
\end{figure}

For the temporal cosmic ray correction, we create a median combined image of the time-series and follow each pixel in the 2D spectral images in time, iteratively replacing 4$\sigma$ outliers with the median pixel value. To remove cosmic rays spatially, we replace outlier pixels within a 2D spectrum with the median value of the surrounding pixels. We can apply the temporal and spatial cosmic ray correction using the custom routines {$\tt remove\_cosmic\_rays\_time$} and {$\tt remove\_cosmic\_rays\_space$} from the {$\tt g280\_transit\_tools$} module. These functions return the cosmic-ray-corrected data, as well as masks for the corrected pixels. The calibrated 2D corrected image compared to the background-subtracted and cosmic-ray-corrected image is shown in Figure \ref{fig:cr_corr}.

\subsection{Embedding Subarrays}
\label{sec:embed_sub}

\begin{figure}[!ht]
\begin{centering}
\includegraphics[width=0.8\textwidth]{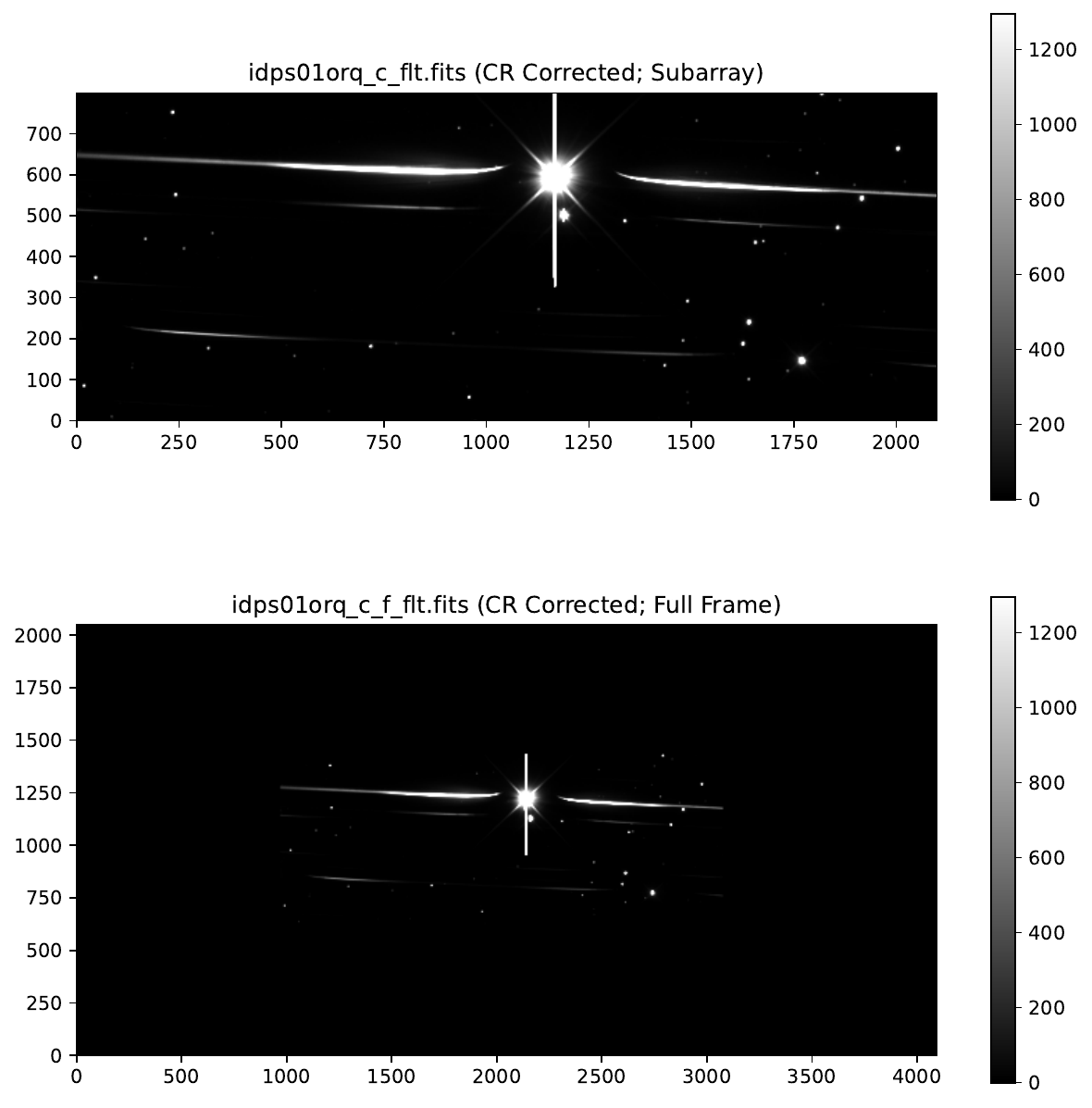}
\caption{The cleaned, cosmic-ray-corrected subarray image (top) compared to the cleaned and embedded full-frame image (bottom).} 
\label{fig:emdbed_sub}   
\end{centering}
\end{figure}

Since the G280 spectra are taken as subarrays on UVIS2, we need to embed this subarray into the full frame in order to proceed with the spectral extraction. To embed the subarrays into the full frame, we use the custom {$\tt embedsub\_uvis$} routine in the {$\tt g280\_transit\_tools$} module, which is a derivative of the {$\tt wfc3tools.embedsub$}\footnote{embedsub: \url{https://wfc3tools.readthedocs.io/en/latest/wfc3tools/embedsub.html\#embedsub}} routine. The custom G280 function that we implement here uses the {$\tt NAXIS$} and {$\tt LTV$} header keywords from the {$\tt flt$} files to convert to the full-frame coordinates when embedding the subarray. This routine circumvents the need to download the {$\tt spt$} files from MAST and use keywords from the {$\tt spt$} headers, as is required for {$\tt wfc3tools.embedsub$}. An example cleaned subarray image compared to an embedded full-frame image is shown in Figure \ref{fig:emdbed_sub}.

\subsection{Spectral Trace Fitting}
\label{sec:trace_fitting}

To fit the spectral trace, we use the custom {$\tt fit\_spectral\_trace$} function from the {$\tt g280\_transit\_tools$} module. This function uses the {$\tt GRISMCONF$} module\footnote{GRISMCONF: \url{https://github.com/npirzkal/GRISMCONF}}, which implements the grism configuration detailed in \textcite{pirzkal2017}. To run the trace fitting function, users must first download the required {$\tt GRISMCONF$} configuration files\footnote{GRISMCONF configuration files: \url{ https://github.com/npirzkal/GRISM_WFC3}}. The trace fitting function returns the $x$ and $y$ spectral traces, the wavelength solution, and the sensitivity function for the specified spectral order. 

\begin{figure}[!ht]
\begin{centering}
\includegraphics[width=0.8\textwidth]{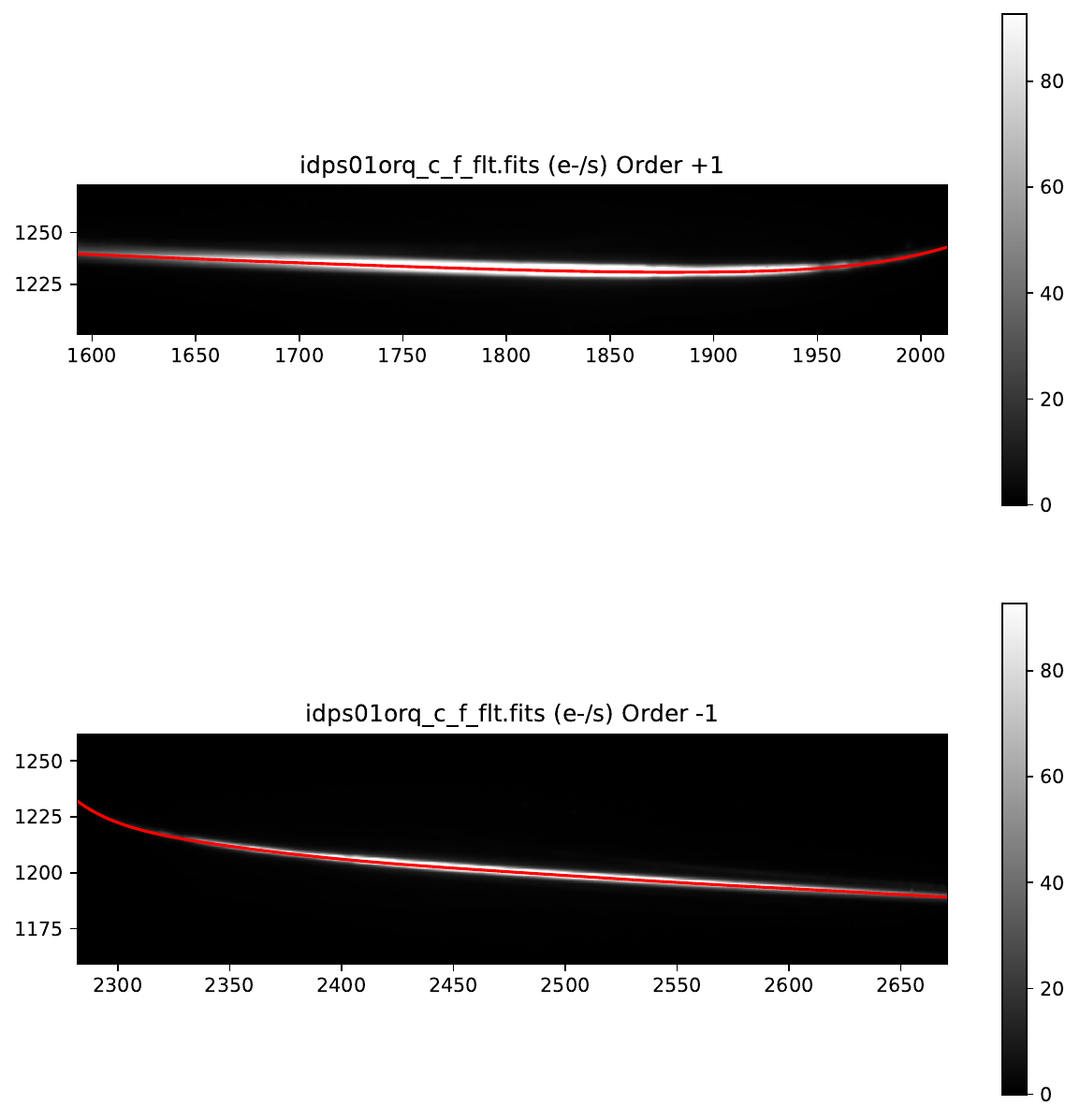}
\caption{The fitted spectral trace (red) for the $+1$ (top) and $-1$ (bottom) G280 orders. } 
\label{fig:trace_fit}   
\end{centering}
\end{figure}

In this tutorial, we extract the $+1$ and $-1$ spectral orders for one of the cleaned full-frame images for the full wavelength range of the G280 grism (0.2--0.8\,$\mu$m), as shown in Figure \ref{fig:trace_fit}. The traces for higher spectral orders (e.g., $\pm$2, $\pm$3, etc.) can also be extracted using this function, and must be specified with the \texttt{order} argument in {$\tt fit\_spectral\_trace$}. 

\subsection{Spectral Extraction}
\label{sec:spec_extract}

\begin{figure}[!ht]
\begin{centering}
\includegraphics[width=0.8\textwidth]{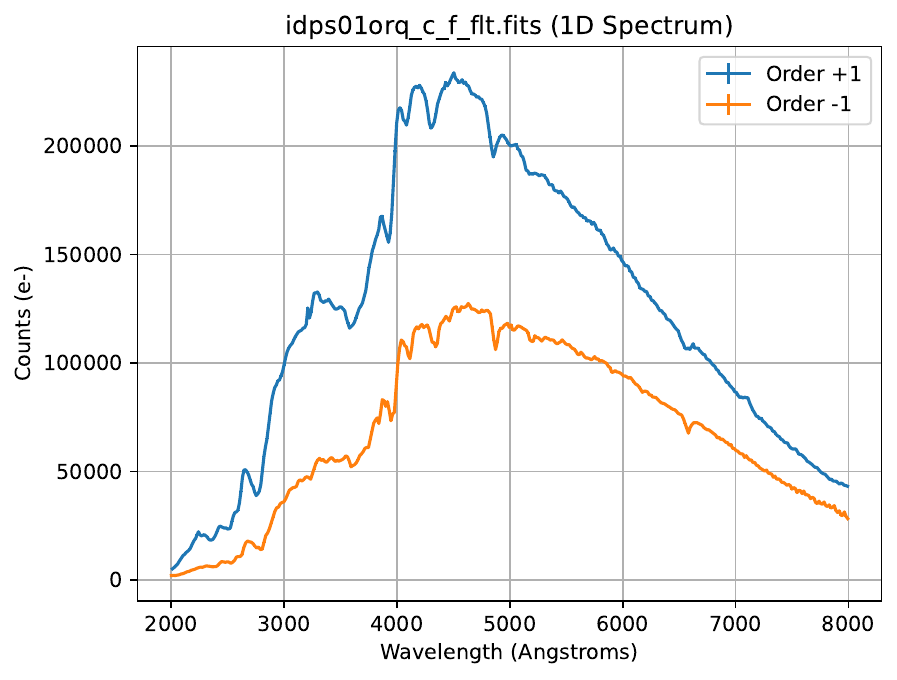}
\caption{Example extracted 1D stellar spectra for the $+1$ (blue) and $-1$ (orange) spectral orders. The $-1$ spectral order has a factor of $\sim$2 fewer counts compared to the $+1$ order.} 
\label{fig:extracted_spec}   
\end{centering}
\end{figure}

Once the traces are fitted, we convert the 2D images to 1D spectra by summing the pixel values for each column along the dispersion direction within an aperture window centered on the spectral trace. This 1D spectral extraction can be performed using the {$\tt extract\_spectrum$} function from the {$\tt g280\_transit\_tools$} module. To run the spectral extraction function, the user must specify the path to the cleaned full-frame grism 2D image, the $x$ trace (outputted by the trace-fitting function), and the aperture window minimum and maximum (in pixels relative to the origin of UVIS2). The aperture window can be refined and optimized later on in the analysis to minimize the root mean square (RMS) scatter in the resulting white light curves. Figure \ref{fig:extracted_spec} shows an example extracted 1D stellar spectrum for the $\pm$1 spectral orders. 

\newpage
\subsection{Extracting 1D Time-Series Stellar Spectra}
\label{sec:time_series}

\begin{figure}[!ht]
\begin{centering}
\includegraphics[width=1.1\textwidth]{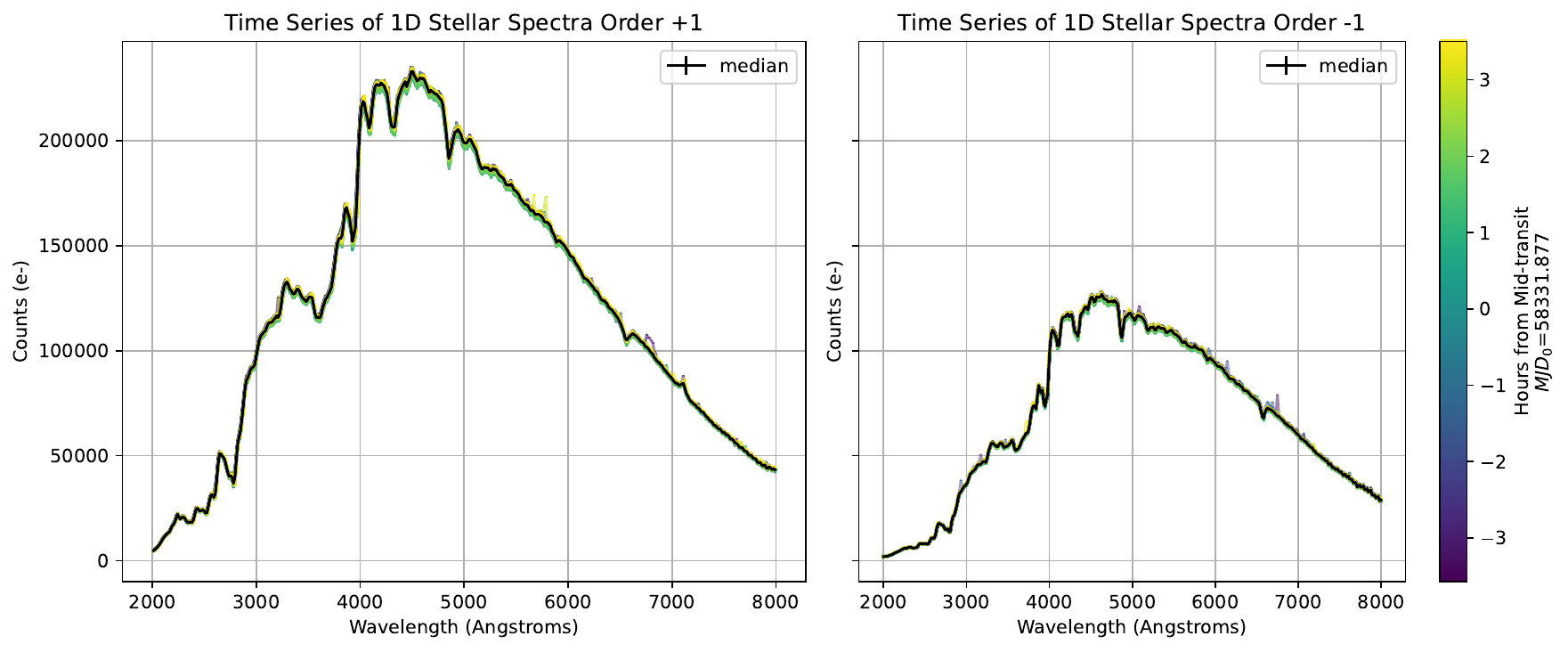}
\caption{The extracted time-series stellar spectra for the $+1$ (left) and $-1$ (right) spectral orders for the transit of HAT-P-41b. The color bar denotes time from mid-transit, with the median spectrum of the time-series shown in black.} 
\label{fig:time_series}   
\end{centering}
\end{figure}

To extract the full time-series of 1D stellar spectra, we can run the trace fitting and spectral extraction routines detailed in Sections \ref{sec:trace_fitting} and \ref{sec:spec_extract}, respectively, for all of the full-frame images. Using the spectral trace fitting parameters, we perform the spectral extraction for all of the full-frame images by running {$\tt g280\_transit\_tools.extract\_spectrum$} on the full time-series, as shown in Figure \ref{fig:time_series}. We can see that there are small offsets in the 1D stellar spectra over the course of the transit observation, which arise from small changes in the position of the spectral trace. 
While we do not perform a correction for these shifts in this tutorial, it is common practice to measure and correct for these shifts in the $x$ and $y$ pixel position on the detector in publication-level scientific analyses (e.g., \textcite{wakeford2020,boehm2024}) by aligning or cross-correlating the time-series before extracting the transit light curves.   

\subsection{Generating Transit Light Curves}
\label{sec:lcs}

\begin{figure}[!ht]
\begin{centering}
\includegraphics[width=0.8\textwidth]{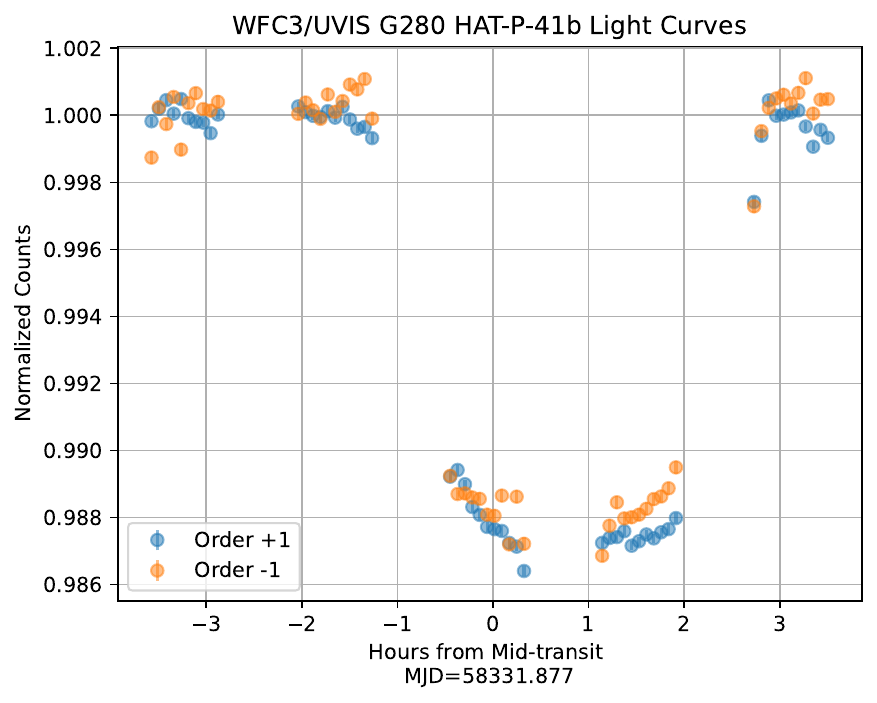}
\caption{The extracted broadband (0.2--0.8\,$\mu$m) transit light curves of HAT-P-41b for the $+1$ (blue) and $-1$ (orange) spectral orders.} 
\label{fig:wlc}   
\end{centering}
\end{figure}

With 1D time-series stellar spectra, we can generate the broadband (i.e., white light) light curves, which are extracted across the full wavelength range of the G280 grism. We generate the light curve using the {$\tt g280\_transit\_tools.make\_light\_curve$} function. The function parameters include the wavelength solution of the order, the time series counts of the order, as well as the minimum and maximum wavelengths over which to extract the transit light curve. 	
The raw white light curves for the $+1$ and $-1$ orders of HAT-P-41b are shown in Figure \ref{fig:wlc}. We can also use the above function to generate the spectroscopic light curves by changing the values of the minimum and maxiumum wavelength parameters, {$\tt wl\_min$} and {$\tt wl\_max $}. 

\section{Summary}
\label{sec:summary}

In this report, we describe a Jupyter notebook for analyzing exoplanet transits observed with the WFC3/UVIS G280 grism. The tutorial details how to download data from MAST, apply a background subtraction and cosmic ray correction for the time-series, perform a trace fitting and spectral extraction for G280 data, and generate broadband and spectroscopic transit light curves. With the modules provided in this tutorial, users should be able to assemble the transmission spectrum for any G280 exoplanet time-series observations and for any of the (positive and negative) spectral orders. With the end products of this notebook -- white light and spectroscopic light curves -- users can use any of the publicly available transit light curve fitting codes (e.g., {$\tt PacMan$}\footnote{PacMan: \url{https://pacmandocs.readthedocs.io/en/latest/}}, {$\tt WFC3$}\footnote{WFC3: \url{https://github.com/kevin218/WFC3}}, or {$\tt Eureka!$}\footnote{Eureka!: \url{https://eurekadocs.readthedocs.io/en/latest/}}) to measure transit depth as a function of wavelength.

\section*{Acknowledgments}

The authors would to thank Peter McCullough and Mariarosa Marinelli for their detailed feedback, which helped to improve this report. We extend our gratitude to all those in the Data Management Division, especially Clare Shanahan, for their careful code review. The authors would also like to thank the editor, Dr. Joel Green, for his review. 

\printbibliography

@ARTICLE{wakeford2020,
       author = {{Wakeford}, H.~R. and {Sing}, D.~K. and {Stevenson}, K.~B. and {Lewis}, N.~K. and {Pirzkal}, N. and {Wilson}, T.~J. and {Goyal}, J. and {Kataria}, T. and {Mikal-Evans}, T. and {Nikolov}, N. and {Spake}, J.},
        title = "{Into the UV: A Precise Transmission Spectrum of HAT-P-41b Using Hubble's WFC3/UVIS G280 Grism}",
      journal = {AJ},
         year = 2020,
        month = may,
       volume = {159},
       number = {5},
          eid = {204},
        pages = {204},
          doi = {10.3847/1538-3881/ab7b78},
archivePrefix = {arXiv},
       eprint = {2003.00536},
 primaryClass = {astro-ph.EP},
       adsurl = {https://ui.adsabs.harvard.edu/abs/2020AJ....159..204W}
}

@ARTICLE{zahnle2009,
       author = {{Zahnle}, K. and {Marley}, M.~S. and {Freedman}, R.~S. and {Lodders}, K. and {Fortney}, J.~J.},
        title = "{Atmospheric Sulfur Photochemistry on Hot Jupiters}",
      journal = {ApJL},
         year = 2009,
        month = aug,
       volume = {701},
       number = {1},
        pages = {L20-L24},
          doi = {10.1088/0004-637X/701/1/L20},
archivePrefix = {arXiv},
       eprint = {0903.1663},
 primaryClass = {astro-ph.EP},
       adsurl = {https://ui.adsabs.harvard.edu/abs/2009ApJ...701L..20Z}
}

@ARTICLE{gao2021,
       author = {{Gao}, Peter and {Wakeford}, Hannah R. and {Moran}, Sarah E. and {Parmentier}, Vivien},
        title = "{Aerosols in Exoplanet Atmospheres}",
      journal = {Journal of Geophysical Research (Planets)},
         year = 2021,
        month = apr,
       volume = {126},
       number = {4},
          eid = {e06655},
        pages = {e06655},
          doi = {10.1029/2020JE006655},
archivePrefix = {arXiv},
       eprint = {2102.03480},
 primaryClass = {astro-ph.EP},
       adsurl = {https://ui.adsabs.harvard.edu/abs/2021JGRE..12606655G}
}

@ARTICLE{tsai2021,
       author = {{Tsai}, Shang-Min and {Malik}, Matej and {Kitzmann}, Daniel and {Lyons}, James R. and {Fateev}, Alexander and {Lee}, Elspeth and {Heng}, Kevin},
        title = "{A Comparative Study of Atmospheric Chemistry with VULCAN}",
      journal = {ApJ},
         year = 2021,
        month = dec,
       volume = {923},
       number = {2},
          eid = {264},
        pages = {264},
          doi = {10.3847/1538-4357/ac29bc},
archivePrefix = {arXiv},
       eprint = {2108.01790},
 primaryClass = {astro-ph.EP},
       adsurl = {https://ui.adsabs.harvard.edu/abs/2021ApJ...923..264T}
}

@ARTICLE{lothringer2022,
       author = {{Lothringer}, Joshua D. and {Sing}, David K. and {Rustamkulov}, Zafar and {Wakeford}, Hannah R. and {Stevenson}, Kevin B. and {Nikolov}, Nikolay and {Lavvas}, Panayotis and {Spake}, Jessica J. and {Winch}, Autumn T.},
        title = "{UV absorption by silicate cloud precursors in ultra-hot Jupiter WASP-178b}",
      journal = {Nature},
         year = 2022,
        month = apr,
       volume = {604},
       number = {7904},
        pages = {49-52},
          doi = {10.1038/s41586-022-04453-2},
archivePrefix = {arXiv},
       eprint = {2204.03639},
 primaryClass = {astro-ph.EP},
       adsurl = {https://ui.adsabs.harvard.edu/abs/2022Natur.604...49L}
}

@ARTICLE{boehm2024,
       author = {{Boehm}, V.~A. and {Lewis}, N.~K. and {Fairman}, C.~E. and {Moran}, S.~E. and {Gasc{\'o}n}, C. and {Wakeford}, H.~R. and {Alam}, M.~K. and {Alderson}, L. and {Barstow}, J. and {Batalha}, N.~E. and {Grant}, D. and {L{\'o}pez-Morales}, M. and {MacDonald}, R.~J. and {Marley}, Mark S. and {Ohno}, K.},
        title = "{The HUSTLE Program: The UV to Near-infrared HST WFC3/UVIS G280 Transmission Spectrum of WASP-127b}",
      journal = {AJ},
         year = 2025,
        month = jan,
       volume = {169},
       number = {1},
          eid = {23},
        pages = {23},
          doi = {10.3847/1538-3881/ad8dde},
archivePrefix = {arXiv},
       eprint = {2410.17368},
 primaryClass = {astro-ph.EP},
       adsurl = {https://ui.adsabs.harvard.edu/abs/2025AJ....169...23B}
}

@MISC{pagul2023,
       author = {{Pagul}, A. and {Ryan}, R. and {Kuhn}, B. and {Som}, D.},
        title = "{The WFC3/UVIS G280 Grism Sky}",
 howpublished = {Instrument Science Report WFC3 2023-06, 11 pages},
         year = 2023,
        month = sep,
        pages = {6},
       adsurl = {https://ui.adsabs.harvard.edu/abs/2023wfc..rept....6P}
}

@MISC{pirzkal2017,
       author = {{Pirzkal}, Norbert and {Ryan}, Russell},
        title = "{A more generalized coordinate transformation approach for grisms}",
 howpublished = {Instrument Science Report WFC3 2017-01 (v.1), 9 pages},
         year = 2017,
        month = jan,
        pages = {1},
       adsurl = {https://ui.adsabs.harvard.edu/abs/2017wfc..rept....1P}
}

@ARTICLE{Ginsburg19,
       author = {{Ginsburg}, Adam and {Sip{\H{o}}cz}, Brigitta M. and {Brasseur}, C.~E. and {Cowperthwaite}, Philip S. and {Craig}, Matthew W. and {Deil}, Christoph and {Guillochon}, James and {Guzman}, Giannina and {Liedtke}, Simon and {Lian Lim}, Pey and {Lockhart}, Kelly E. and {Mommert}, Michael and {Morris}, Brett M. and {Norman}, Henrik and {Parikh}, Madhura and {Persson}, Magnus V. and {Robitaille}, Thomas P. and {Segovia}, Juan-Carlos and {Singer}, Leo P. and {Tollerud}, Erik J. and {de Val-Borro}, Miguel and {Valtchanov}, Ivan and {Woillez}, Julien and {Astroquery Collaboration} and {a subset of astropy Collaboration}},
        title = "{astroquery: An Astronomical Web-querying Package in Python}",
      journal = {AJ},
         year = 2019,
        month = mar,
       volume = {157},
       number = {3},
          eid = {98},
        pages = {98},
          doi = {10.3847/1538-3881/aafc33},
archivePrefix = {arXiv},
       eprint = {1901.04520},
 primaryClass = {astro-ph.IM},
       adsurl = {https://ui.adsabs.harvard.edu/abs/2019AJ....157...98G}
}

\end{document}